# Measurement of g-factors both ground and excited optical states in zero dc magnetic field by photon echo method

**V N Lisin, A M Shegeda, V V Samartsev**

*Zavoisky Physical–Technical Institute, Russian Academy of Sciences, Kazan, 420029 Russia*

*e-mail: vlisin@kfti.knc.ru*

New scheme of definition of g-factors as ground as excited optical states of a paramagnetic ion in zero external constant magnetic field has been proposed and experimentally realized in optical systems in which Zeeman Effect is manifested. A pulse of a weak magnetic field leads to occurrence of relative phase shifts of the excited dipoles and, as consequence, to modulation of a photon echo waveform if magnetic pulse (MP) overlaps in time with echo-pulse. The modulation periods of the waveform depend on polarization of the laser light, which excites the photon echo. The values of these periods for σ- and π-laser light polarization have been measured and then the g-factors of the ground $^4I_{15/2}$ and excited $^4F_{9/2}$ states of the $Er^{3+}$ ion in the $LuLiF_4$ and the $YLiF_4$ matrices have been determined. Values of the g-factors have been compared with the known literary data.

**1. Introduction**

Paper purpose is to measure by the photon echo method the ground and excited states g-factor values of paramagnetic ion $Er^{3+}$ in $LuLiF_4$ and $YLiF_4$ in magnetic fields, which are much less then fields used in EPR. The $YLiF_4$ matrice has been used for control.

Photon echo is coherent radiation of medium in form of short pulse, caused by restoration of phase of separate radiators after the change of a sign on relative frequency of radiators. It is essential that each excited radiator give the identical contribution to the resulting dipole moment. If pulsed perturbation shifts the transition frequency of radiators for the same amount, echo intensity is not changed. In systems in which the Zeeman Effect is manifested a pulsed magnetic field shifts the optical frequencies of two groups of paramagnetic ions with spin projection in ground state ±1/2 on the different values (as you can see from figure 1). Relative phase $\varphi$ of these dipole groups of is not equal to zero as result. Echo intensity oscillates versus phase [1, 2]:

$$I(t) = I_0(t)(1+\cos\varphi(t))/2, \quad \varphi(t) = \int_{t_0}^{t} Z(t')dt', \tag{1}$$

where $I_0(t)$ is intensity of an echo when a MP off, $t_0$ is time of the beginning of action of a MP, $Z$ is Zeeman splitting of optical line.

If MP does not overlap in time with echo pulse and laser pulses then phase $\varphi$ is not depended on time. As result the echo intensity oscillates versus phase without of waveform changing. Zeeman splitting equal 10 MHz was observed in [1]. That is much less than the inhomogeneous line width equal 600 MHz in YLiF$_4$: Er$^{3+}$.

When MP overlaps in time with echo pulse then phase becomes time-dependent. Expressions for dipole moment and intensity of echo formally do not vary (because perturbation is diagonal). Echo time form is changed. If relative phase $\varphi$ is larger then $3\pi$ then the time form of an echo is modulated. There is enough one waveform [2] to determine modulation period $T$ and therefore Zeeman splitting:

$$\varphi(t+T) - \varphi(t) = 2\pi = \int_{t}^{t+T} Z(t')dt'. \tag{2}$$

Selection rules of the transitions for $\pi$- and $\sigma$- polarizations of the laser are different as you can see in figure 1.

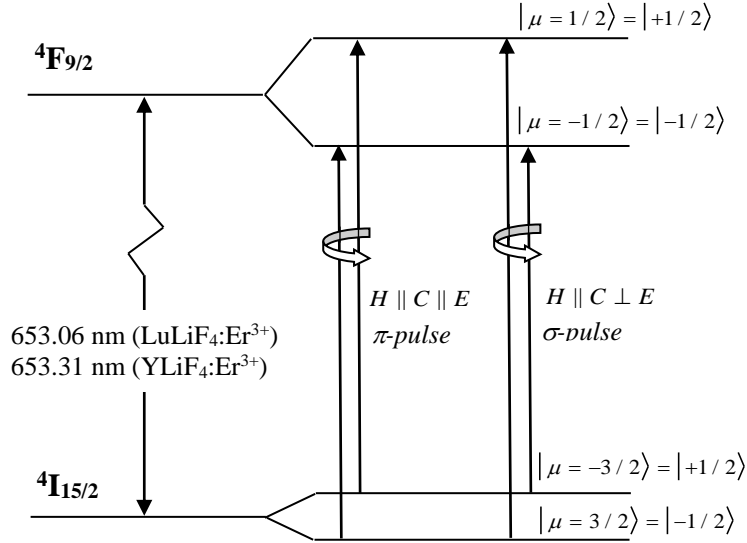

Figure 1. Energy levels Er$^{3+}$ in YLiF4 and LuLiF4 ($\mathbf{H} \parallel \mathbf{C}$). Also transition and transition wavelength are shown (the arrows indicate the transitions for $\sigma$- and $\pi$- polarizations of the laser.). $\mathbf{C}$ is crystal axes, $\mathbf{H}$ is magnetic field and $\mathbf{E}$ is laser electrical field. $\mathbf{E} \parallel \mathbf{C}$ – $\pi$- polarization, $\mathbf{E} \perp \mathbf{C}$ – $\sigma$- polarization. $\mu$ is crystal quantum number, $|\pm 1/2\rangle$ indicate the Er$^{3+}$ ion states which spin projection on axis $\mathbf{C}$ are $\pm 1/2$.

Zeeman splitting $\pi$- and $\sigma$- optical lines are different also:

$$\begin{aligned} Z_{\sigma,\pi}(t) &= (g' \pm g)\beta H(t)/\hbar, \\ g &= g_{\parallel}(^4I_{15/2}), \quad g' = g_{\parallel}(^4F_{9/2}). \end{aligned} \tag{3}$$

Here $g$ and $g'$ - components the ground and excited states g-tensor, parallel to a crystal axis $C$, $H$ - amplitude of a pulse magnetic field, signs plus and minus in the expression corresponds to Zeeman splitting $Z_\sigma$ and $Z_\pi$ of frequency of the transition excited by $\sigma$- and $\pi$- polarized light. Then the ground and excited states g-factors can be determined, use (1), (2) and (3), from measurement of waveform modulation periods of echo excited by $\pi$- and $\sigma$- polarized lasers (two equations and two unknown values).

Let's pass to consideration of experimental technics

## 2. Experimental conditions

We described the details of the experiment in [1, 2]. For this reason, we present only the main moments and some differences. We observed the photon echo in the reverse mode on the

$^4I_{15/2}$ -> $^4F_{9/2}$ transition of the $Er^{3+}$ ion at temperature of 2 K. Laser pulses have been generated by an oxazine-17 dye laser. Duration is 20 ns, spectral FWHM is 0.15 Å = 9 GHz. Delay time between the laser pulses is 35 ns. Polarization of the laser pulses can be set parallel to the axis $C$ ($\pi$-polarization) or perpendicular to the $C$ ($\sigma$-polarization). The photon echo signals have been detected on a photomultiplier and fed to a Tektronix TDS 2022 digital oscilloscope. The rise time of the oscilloscope TDS-2020 is ~ 2.1 ns. All experimental measurements have been controlled with LabVIEW software. The experiment setup shown in figure 2.

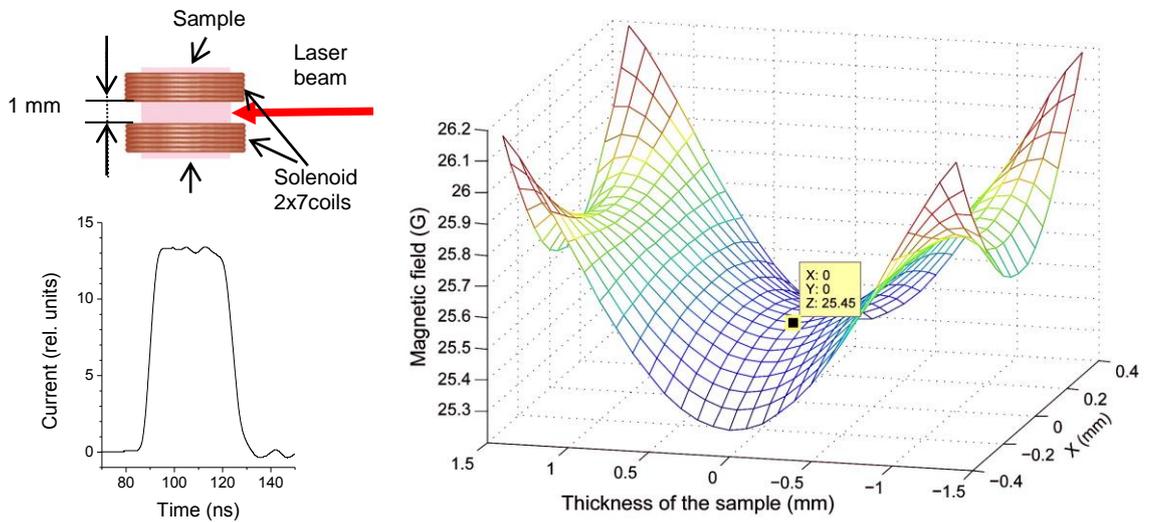

Figure 2. Geometry of the experiment shown on the left in the top picture. In the down the current pulse flowing through the solenoid coil is shown. The **C** axis of the sample is parallel to the solenoid axis. The right shows the calculated component of the magnetic field $H$ parallel to the C axis of the sample in the solenoid with a diameter of 5.1 mm at the current in the solenoid coils of 1 A. The variation of $H$ along the solenoid axis within a diameter of 0.2 mm of the optically excited region is shown on the $X$-axis; the variation of $H$ along the optically excited region of the $LiLuF_4:Er^{3+}$ sample is shown on the Y-axis. $H$ = 25.45 G in the center of the solenoid (x = 0, y = 0).

The studied sample has been placed in a solenoid consisting of two coaxial coils each having seven turns of PEV-0.18 copper wire (the diameter of a wire with isolation was 0.2 mm). The coils were connected in parallel in order to decrease the solenoid inductivity and thus to improve the current pulse shape. The *C* axis of the sample was parallel to the solenoid axis. The internal solenoid diameter for the $LuLiF_4:Er^{3+}$ and $YLiF_4:Er^{3+}$ samples were 4.9 and 6.2 mm, respectively. Consistently with solenoids the resistors included having resistance at helium temperatures 53 Ohm and 54.3 Ohm for samples $LuLiF_4:Er^{3+}$ and $YLiF_4:Er^{3+}$ accordingly.

Current pulses formed by a generator with avalanche transistors have been applied to the solenoid synchronously with the laser pulses. The half max current pulse duration $\tau_H$ is equal 34 ns and upper flat part duration is 25 ns as you can see in figure 2. The maximum amplitude of a current is equal 0.764A and 0.607A through solenoids with samples $LuLiF_4:Er^{3+}$ and $YLiF_4:Er^{3+}$ accordingly.

The pulsed magnetic field arising in gap between solenoid coils has been calculated using the Biot–Savart law. The dependence of resistance on temperature as well as the wire insulation thickness has been taken into account: solenoid diameter for the $LuLiF_4:Er^{3+}$ and $YLiF_4:Er^{3+}$ samples were 5.1 and 6.4 mm when calculating. The calculated values agrees well with measured with a Hall probe in gap center in the case of dc current through the solenoid at room temperature. Therefore, the amplitude and shape of the MP are determined by the amplitude and shape of the current pulse through the solenoid.

Calculation shows that value of components of the magnetic field, a parallel axis *C*, in the center of the solenoid in diameter of 5.1 mm equally $H = 25.45$ G and for the solenoid in diameter of 6.4 mm $H = 22.16$ G at a current through each coil 1A. The values of a magnetic field calculated for a room temperature in the center of solenoids well enough coincide with measured macroscopic Hall probe at direct current through solenoids. The average value of the field inhomogeneity is determined only by the coil geometry. It is $1.26 \times 10^{-2}$ along the axis *Y* and $-3.12 \times 10^{-4}$ along the axis *X*.

The current amplitude has been varied with a step of 1 dB. However, inductance presence in loading increased an attenuator error, its calibration therefore became. Table 1 shows the relative change in the amplitude of the current pulses with the different settings of the attenuator and the magnetic field values in the gap center (*x*=0, *y*=0) of the solenoids with samples $LuLiF_4:Er^{3+}$ and $YLiF_4:Er^{3+}$:

Table 1.

| dB | 0 | 1 | 2 | 3 | 4 | 5 | 6 | 7 |
|---|---|---|---|---|---|---|---|---|
| Current decay | 1 | 0,883 | 0,776 | 0,684 | 0,604 | 0,536 | 0,473 | 0,419 |
| $H$ (G), $LuLiF_4:Er^{3+}$ | 19.44 | 17.17 | 15.08 | 13.3 | 11.74 | 10.42 | 9,19 | 8,15 |
| $H$ (G), $LuLiF_4:Er^{3+}$ | 13.45 | 11.87 | 10.44 | 9.2 | 8.11 | 7.21 | 6.36 | 5.63 |

## 3. Results of experiment

In figure 3 waveforms of observable echo signals in samples LuLiF$_4$:Er$^{3+}$ and YLiF$_4$:Er$^{3+}$ are shown. You can see also MP waveforms calculated in the gap center ($x = 0$, $y = 0$) of the solenoids.

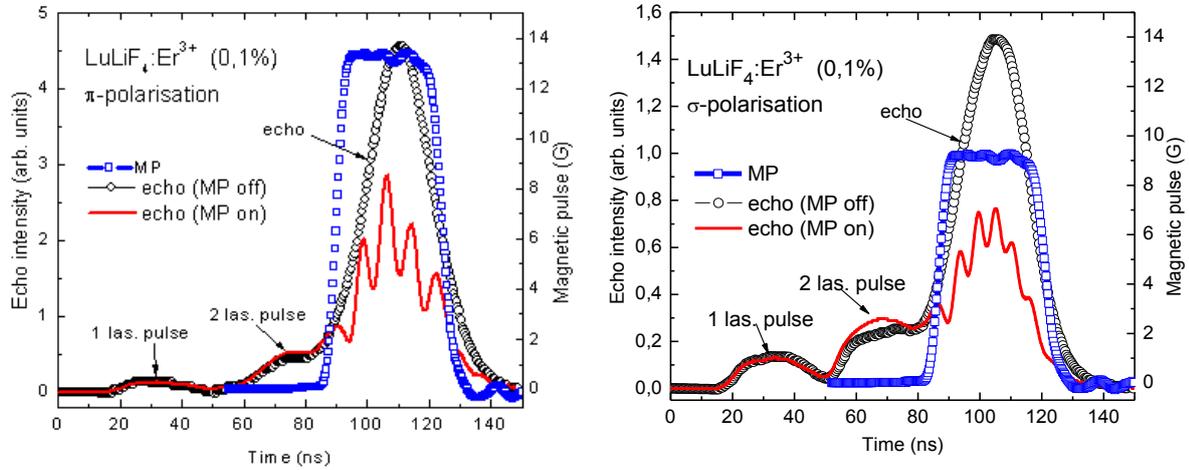

Figure 3. Waveforms of MP and photon echo for laser pulses of different polarizations when the MP on and off. Echo half max duration (MP off) is 28 ns, MP half max duration is 25 ns.

From the waveforms of echo signals observed at various times of the beginning of action of the MP, similar to those shown in figure 3, time intervals between the nearest minima in the echo signals were measured and the average value $T$ of the period has been determined. The magnetic field amplitude $H$ in the region of the top flat time area of current pulse has been calculated in the gap center ($x = 0$, $y = 0$) of the solenoids. The dependence of the inverse average period $1/T$ versus $H$ has been built. Figure 4 shows these dependences for different polarizations of the laser pulses in the YLiF$_4$:Er$^{3+}$ and LuLiF$_4$:Er$^{3+}$. We take into account that echo modulation is absent, i.e. $1/T = 0$ when $H = 0$.

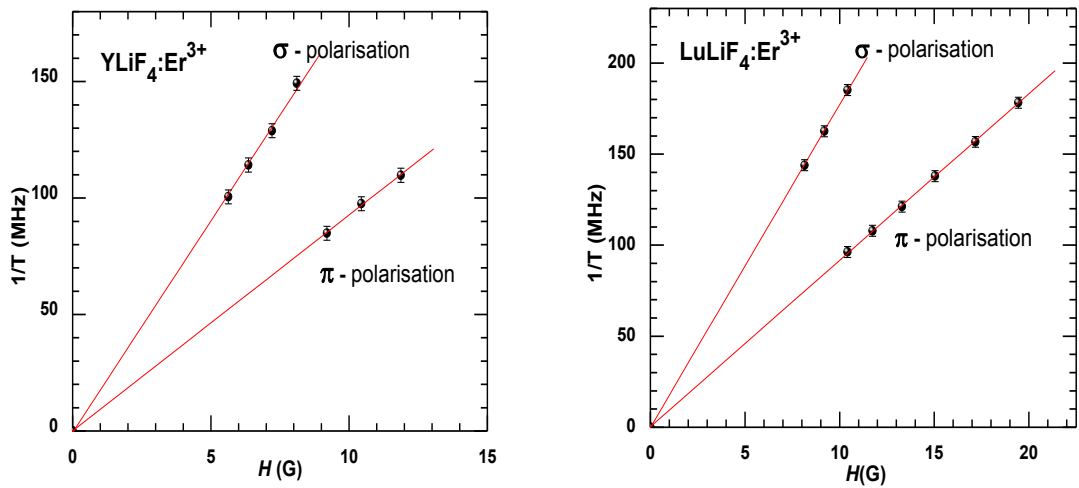

Figure 4. Inverse modulation period of the echo waveform versus $H$. The $H$ is magnetic field amplitude calculated in the gap center ($x = 0$, $y = 0$) of the solenoids for the time in the time region of the top flat of current pulse.

From figure 4 the values of following parameters have been defined

$$\left.\begin{array}{l}\partial(1/T_\sigma)/\partial H\big|_{\exp} = 18.069\,(MHz/G), \\ \partial(1/T_\pi)/\partial H\big|_{\exp} = 9.274\,(MHz/G),\end{array}\right\} YLiF_4 : Er^{3+}$$

$$\left.\begin{array}{l}\partial(1/T_\sigma)/\partial H\big|_{\exp} = 17.716\,(MHz/G), \\ \partial(1/T_\pi)/\partial H\big|_{\exp} = 9.156\,(MHz/G),\end{array}\right\} LuLiF_4 : Er^{3+}$$

(4)

Taking into account that change of a phase during period equal $2\pi$, we do approach

$$\varphi(t+T) - \varphi(t) = 2\pi = \int_t^{t+T} Z(t')dt' \approx ZT,$$

$$Z \approx (g' \pm g)\beta H /(2\pi\hbar T),$$

(5)

where $H$ is MP amplitude at the top flat part. These equations are true, strictly speaking, only on a flat part of a MP. From (5) it is possible to express Zeeman splitting through the inverse period of modulation

$$Z/(2\pi) = (g' \pm g)\beta H /(2\pi\hbar) \approx 1/T \qquad (6)$$

and the sum and a difference of g-factors through experimentally measured parameters

$$\partial(Z_{\sigma,\pi}/(2\pi))/\partial H = (g' \pm g)\beta/(2\pi\hbar) = 1.3996\ (MHz/G)\,(g' \pm g) \approx \partial(1/T_{\sigma,\pi})/\partial H\big|_{\exp}. \qquad (7)$$

The expressions (4) and (7) make it easy to find the difference ($\pi$-polarization) and the sum ($\sigma$-polarization) g-factors and therefore the values of the excited and ground state g-factors:

$$\left.\begin{array}{l}g'+g = 12.658 \\ g'-g = 6.543\end{array}\right\} g = g(^4I_{15/2}) = 3.058,\ g' = g(^4F_{9/2}) = 9.60,\quad LuLiF_4 : Er^{3+},$$

$$\left.\begin{array}{l}g'+g = 12.908 \\ g'-g = 6.622\end{array}\right\} g = g(^4I_{15/2}) = 3.143,\ g' = g(^4F_{9/2}) = 9.765,\ YLiF_4 : Er^{3+}.$$

(8)

For comparison, we give the values obtained by other methods
:

$$\left.\begin{array}{l}g = g(^4I_{15/2}) = 3.09\ [4] \\ g' = g(^4F_{9/2}) = 10.35\ [1],\ 10.27\ [2]\end{array}\right\} LuLiF_4 : Er^{3+}$$

$$\left.\begin{array}{l}g = g(^4I_{15/2}) = 3.14\ [5] \\ g' = g(^4F_{9/2}) = 9.84 \pm 0.25\ [6],\ 9.61\ [7],\ 10.17\ [1],\ 9.88\ [2]\end{array}\right\} YLiF_4 : Er^{3+}$$

(9)

## 4. Discussion of results

There are some errors associated with the measurement error of the current pulse amplitude, as well as a possible error of the oscilloscope time base. We note that the ratio sum of g-factors of the excited and ground state to the difference between them does not depend on the exact value of the magnetic field and the exact value of time interval between the minima but only on their relationship.

This helps to reduce the errors in the determination of, for example, the excited state g- factor $g'$ if the value of $g$ is known. Using (4), we find:

$$\frac{\partial(1/T_\sigma)/\partial H|_{exp}}{\partial(1/T_\pi)/\partial H|_{exp}} = \frac{g'+g}{g'-g} = \begin{cases} 1.95, \text{YLiF}_4:\text{Er}^{3+}, g=3{,}14 \Rightarrow g'=9{,}755, \\ 1.93, \text{LuLiF}_4:\text{Er}^{3+}, g=3{,}09 \Rightarrow g'=9{,}61. \end{cases}$$

The values thus obtained for the g-factors of the excited state is almost identical to the above in (8) results. The accuracy of the calculation of the magnetic field pulse is confirmed once again.

The obtained values of g-factors (8) are in good agreement with the data obtained in [4, 5, 6, 7], but differ substantially from the results [1, 2]. In our previous works [1, 2] the duration of the laser pulses $\tau_L$ was 12 ns, and the duration $\tau_H$ of the MP was about of about 14 ns (shorter than the duration of the echo response). Under such conditions, only two minima have been observed in the echo waveform, because the increase of the amplitude of MPs to increase the number of observed minima would lead to restrictions related to the bandwidth of the oscilloscope, since the rise time of the oscilloscope TDS-2020 is ~ 2.1 ns. The approximation (5) is not true in this case. The accuracy of the measurements has been significantly improved when we have significantly increased the duration of laser ($\tau_L$=12 ns $\Rightarrow$ 20 ns) and magnetic ($\tau_H$=15 ns $\Rightarrow$ 34 ns) pulses. It is important that the duration of the flat part on top of the magnetic pulse $\tau_H$(flat) = 25 ns becomes approximately equal to the echo pulse duration $\tau_{echo}$ = 28 ns. Also the sum and difference of the g-factors values found from $\sigma$- and $\pi$- Zeeman splitting coincide with the known values for the erbium ion in the YLiF$_4$ matrix.

In our opinion, the main reason for the error in the determination of the g-factors, in our case, can be unstable laser pulse shape and, consequently, unstable waveform of the photon echo. In the ideal case, when the initial shape of the echo signal and the shape of the MP close to the square , it is enough a single waveform echo when the MP on to determine the difference ( for $\pi$- polarization) or sum ( for $\sigma$- polarization) g- factors of the ground and excited states. Enough to know the amplitude of the MP and the time between the nearest minima in the resulting echo response. If the shape of the echo signal is completely arbitrary, but repetitive from pulse to pulse , knowing the shape, the value and time $t_0$ when the current pulse on , it is easy to calculate for a given $g$ and $g'$ ( see (1, 3)) the waveform of the resulting echo response . Comparing the period of the modulation of the echo signal with the experiment, we can find the difference (for $\pi$- polarization) or the sum (for $\sigma$- polarization) g- factors. In this case, you have two echo waveforms: with MP off and MP on. If the shape of the echo signal is completely arbitrary and does not repetitive from pulse to pulse, it is impossible to know exactly the shape of the echo when MP off. In this case, to reduce the error we must submit the MP that creates in the echo waveform as much as possible minima and then to calculate the average the time between the nearest minima.

Figure 4 shows that increasing of the MP amplitude has no meaning if the time between the nearest minima becomes less than 5 ns, since the rise time of the oscilloscope TDS-2020 is ~ 2.1 ns. On the other hand, the minimum value of the MP amplitude is limited by the duration of the echo pulse, since it is necessary that at least two minima were distinguishable in the echo waveform.

Obtained in the present work the values of the ground state g-factors at magnetic field amplitude $H$ <20 G are in good agreement with those measured by EPR [4, 5]. The method is free from the weakness of the excited state g-factor measurements at high magnetic fields, when it is necessary to take into account the quadratic magnetic field shift of the optical frequency.

Determination of g-factors by the echo waveform modulation method is significantly less labor intensive than the determination of g-factors for the period of the oscillations [1] the intensity of the echo. To increase accuracy, it is desirable to use longer laser pulses.

Preliminary results had been presented at conferences [8, 9].

## 5. Conclusions

By controlling the relative phase of the excited dipoles with weak perturbation you can significantly change the resultant dipole moment of the system. This can occur in cases where the coherence is important, for example, photon echo [1, 2], Raman heterodyne optical detection [10], radiation propagation [11], and so on. In our study we measured the Zeeman splitting of the ground and excited states and determined magnetic parameters of the paramagnetic ion with an accuracy comparable to the EPR using the known values of the MP amplitude. If the magnetic parameters are known, it is possible to determine the amplitude of the MP, using the relation (6). For example, if the magnetic field is created by the pulse changes of the dipole-dipole interaction of the ion with the environment. This can be used to determine the distance to the centers of the environment by measuring the waveform of the echo modulation periods, not only in the optical range, but also in the EPR and NMR bands, using relations similar to (6).